\documentstyle[aps]{revtex}
\begin{document}
\title{The topology of events}
\author{Vipul Periwal}
\address{Department of Physics,
Princeton University,
Princeton, New Jersey 08544}

\def\al{\alpha}
\def\be{\beta}
\def\eps{\epsilon}
\def\dd{\hbox{d}}
\def\tr{\hbox{tr}}
\def\Tr{\hbox{Tr}}
\maketitle
\begin{abstract}  
A matrix model is constructed to compute characteristic numbers of
the space of  subsets of ${\bf R}^{d} $  with $N$ elements.  This matrix model
is found to be a constrained null dimensional reduction to a point 
of a Yang-Mills
theory  with anticommuting matter  in $d+2$ dimensions.
The constraints and equations of motion  are similar
to those found by Nishino and Sezgin in their $10+2$ dimensional
supersymmetric Yang-Mills equations.  It is conjectured that the $d=10$
topological model is a twisted form of the Nishino-Sezgin model,
dimensionally reduced to a point, due to SO(8) triality.  
Topological matrix models are constructed for non-commutative spaces,
entirely in terms of algebras associated with such spaces.  These 
models exhibit degrees of freedom analogous to massive modes in string 
theory.
\end{abstract} 
An amusing feature of much recent work (reviewed in
\cite{banks,wati,sen}) on matrix models of fundamental
interactions 
is the growing interest in noncommutative geometry, and 
the possibility that  the spacetime continuum of 
common experience is replaced by something described by matrix 
co\"ordinates, first observed in \cite{ed}.

I consider in this letter an {\it a priori} different problem, that 
of calculating some characteristic numbers of the space of finite subsets of 
cardinality $N$ in some manifold, $M.$  I show that, when $M={\bf 
R}^{d},$ this problem is solved by a topological matrix model that 
resembles a dimensional reduction to a point of a Yang-Mills model in
$d+2$ dimensions.  I find that the features of this topological 
matrix model are in close correspondence to the dimensional reduction 
of the Nishino-Sezgin supersymmetric Yang-Mills 
equations\cite{ns,nishino}, which I 
proposed some time ago as a fundamental description of M 
theory\cite{me}.  This leads to the conjecture that precisely when 
$d=10,$ the topological model can be twisted to give rise to the 
(reduced) Nishino-Sezgin model, on account of SO(8) triality.  
Thus, consistent with the spin-statistics theorem, 
only when $d=9+1$ does the topological model possess a sensible 
spacetime interpretation.
If this conjecture is true, non-renormalization properties of 
the reduced supersymmetric Yang-Mills theory would follow for a restricted 
class of observables from the topological invariance of the twisted 
theory.

In this construction that the reduced Yang-Mills form of the action 
appears because the functions on ${\bf R}^{d}$ form a commutative algebra.  It 
is then a simple matter to find a topological model appropriate for 
describing $N$ points in an arbitrary space, in terms of the 
(possibly non-commutative) algebra  associated with the 
space\cite{von}.  Some  features of this algebraic construction 
are taken from Ref.~\cite{thesis}. 
There appear additional degrees of freedom, even in
the case of ${\bf R}^{d},$ that may be analogous to the massive modes 
in string theory.

Lastly, I point out that the appropriate mathematical structure 
underlying these constructions seems to be  a variant of 
operads\cite{cohen}.  Operads appear, of course, 
in conformal field theory\cite{stasheff}
and in string field theory\cite{zwiebach} already, though the  structures that 
appear here seem  distinct.  This paper contains only a 
physical approach to the problem.  I hope to return to the  
mathematical framework elsewhere.

Consider a set with $N$  indistinguishable points, $B_{N}.$  The standard 
algebra of functions to associate with $B_{N}$ is the cross product algebra 
obtained from the Cartesian product of $N$ copies of ${\bf R}$ by the 
action of the symmetry group on $N$ letters.  Another algebra one could
associate with $B_{N}$, is the algebra of $N\times N$ Hermitian 
matrices, $H_{N},$ up to unitary equivalence---the eigenvalues of
the matrices, up to permutations, are the analogues of the values 
of a function on the $N$ points.   This is the algebra 
that arises naturally in Witten's description\cite{ed} of D-instantons.
D-instantons have been studied in Ref.~\cite{gg}. 

\def\rd{{\bf R}^{d}}
\def\cd{{\cal C}({\bf R})^{\otimes d}}
\def\ga{\gamma}
Now consider the space of maps from $B_{N}$ to ${\bf R}^{d}.$  In 
terms of co\"ordinates on $\rd,$ each map  
$x^{i}:\{1,\ldots,N\} \rightarrow \{x^{i}(1),\ldots,x^{i}(N)\},$ so 
this space 
is equivalently described as the space of subsets of $\rd$ with $N$ 
elements. Such 
maps give rise to obvious homomorphisms from the 
continuous functions on ${\bf R}^{d},$ ${\cal C}({\bf R}^{d})
\equiv {\cal C}({\bf R})^{\otimes d} $
to $
H_{N},$ by pullback: Every element $f\in\cd$ gets mapped to
$x^{*}f(\{1,\ldots,N\}) \equiv {\rm 
diag}\{f(x(1)),f(x(2)),\ldots,f(x(N))\},$
up to equivalence by unitary conjugation.

\def\ps#1{\psi^{{#1}}}
To start with, we will suppose that an element of 
$Hom(\cd, H_{N})$ is specified by its action on the generators of
$\cd,$ which we take to be the co\"ordinate functions, $x^{i}.$
Thus, an element of $Hom(\cd, H_{N})$ is specified by $d$ commuting
elements of $H_{N},$ $X^{i}.$
We want to compute topological invariants of $Hom(\cd, H_{N}).$  To 
this end, we want to construct a measure to integrate over $d$ 
arbitrary
elements of $H_{N},$ which is concentrated on configurations of $d$ 
commuting matrices.  We 
want the measure to be invariant under unitary conjugation, in order 
that we describe only unordered sets. 

There is a standard procedure 
for constructing such measures, extracted from Witten's work\cite{edtop} 
on a quantum field-theoretic interpretation of the Donaldson
invariants.  One introduces a set of $d$ matrices with anticommuting 
entries, 
which we label $\psi^{j}.$  Further we define an anticommuting
operator $s$ acting 
on $X^{i}, \psi^{j}:$
\begin{equation}
sX^{i} = \ps i, \qquad s\ps i = i[\phi,X^{i}], \qquad  s\phi = 0
\label{sX}
\end{equation}
where $\phi$ is a matrix with commuting entries.  
We introduce a set of matrices with anticommuting entries, $b^{ij}$ 
(antisymmetric in $i,j$)   
and $L,$ and a matrix with commuting entries, $K,$
on which $s$ acts as
\begin{equation}
sb^{ij}=i[X^{i},X^{j}], \qquad sK = L,\qquad sL = 
i[\phi,K].
\label{sb}
\end{equation}
For orientation, note that $s$ is the equivariant BRS operator for a topological
shift symmetry, $b^{ij}$ will play the r\^ole of the antighost for
the constraint of interest $[X^{i},X^{j}]=0$, $\ps 
i$ is the ghost for the topological shift symmetry, and $K$ is the
antighost for the constraint $[X_{i},\ps i]=0.$  The latter 
constraint is standard in constructions of topological field theories, 
and ensures that zero-modes of $\ps i$ are in 1--1 correspondence 
with moduli deformations.  To see this, note that if $\{X^{i}\}$ is
a solution of 
$[X^{i},X^{j}]=0,$ $\{X^{i} + i[X^{i},E]\}$ is also a solution, for 
any matrix $E.$  Thus 
the $\psi$ equations of motion will automatically have 
solutions of the form $i\eps[X^{i},E],$ where $\eps$ is an anticommuting 
number. If we demand $[X_{i},\psi^{i}]=0,$  such solutions are 
eliminated, leaving only the desired solutions.
    
\def\ps#1{b^{+{#1}}}
Relabel the matrices as follows: Let $I,J \in \{1,\ldots,d,+,-\},$ and
define $\ps I \equiv \psi^{I},$ with $ \psi^{-}\equiv 
L.$  By antisymmetry, $\ps +=0,$ and   we constrain $b^{-j}=0.$  
Further, define $X^{+}\equiv\phi,$ 
and
$X^{-}\equiv K.$  We see now that the action of $s$ given in 
eq.'s~\ref{sX},\ref{sb} may be written 
more succinctly as 
\begin{equation}
sX^{I}\equiv b^{+I},\qquad sb^{+I} = i[X^{+},X^{I}], \qquad 
sb^{ij}=i[X^{i},X^{j}].
\label{sdplus2}
\end{equation}
While the matrices are naturally organized in this sort of way, there 
is clearly no $d+2$ dimensional Lorentz-like symmetry in the action 
of $s,$ since the $+$ index appears explicitly in eq.~\ref{sdplus2}. 
With the usual 
ghost numbers, we see that there is a simple rule:
$gh(X^{i})=0, gh(X^{\pm}) = \pm2,$ and $gh(b^{ij})=-1, 
gh(\ps i)=+1, gh(\ps -) = -1.$  Thus an upper $\pm$ 
index corresponds to an additional ghost number $\pm2.$   It seems 
natural to think of this 
ghost number as the analogue of Ramond-Ramond charge.  It would be 
interesting to relate this to the work of Berkovits\cite{nathan}.

We will need a metric for raising and lowering indices.  Let us 
suppose that the metric on $\rd$ is Minkowskian of some signature 
$(p,d-p)$---different choices are irrelevant for the topological 
theory, but are relevant for the physical theory (see below).  We extend the 
inner product to include the $\pm$ indices so that 
$(v^{i},v^{+},v^{-})\cdot (w^{i},w^{+},w^{-}) = v^{i}w_{i} 
+v^{+}w^{-}+v^{-}w^{+}.$  Now 
consider the gauge fermion
\begin{equation}
\Psi = i\tr\left( b_{ij}[X^{i},X^{j}] + 2 X^{-}[X_{i},\ps i]\right)= i\tr 
\left(b_{ij}[X^{i},X^{j}] + 2 \ps i[X^{-},X_{i}] \right) .
\label{gaugefermion}
\end{equation}
We have 
\[ s\Psi = i\tr\bigg( i[X_{i},X_{j}][X^{i},X^{j}] +2i[X^{-},X_{i}][X^{+},X^{i}]
+b_{ij}([X^{j},\ps i] - 
[X^{i},\ps j ]) + 2 \ps - [X_{i},\ps i] -2b^{+}{}_{i}[X^{-},\ps i]
  \bigg).\] 
Let $\Delta S \equiv {i\over 2}\tr (b_{ij}[X^{+},b^{ij}]).$
Define the complete action $S\equiv s\Psi +\Delta S.$
All    terms with anticommuting fields have an 
upper $+$ index, the analogue of a chiral projection for fermion terms. 
Note that $s^{2}\cdot = i[X^{+},\cdot],$ and computing 
$sS= s^{2}\Psi + s\Delta S,$ we find
$sS=0,$ using the $b_{ij}$
equation of motion.  
$s$ is therefore a scalar supersymmetry, squaring 
to a bosonic field-dependent infinitesimal U($N$) transformation, the 
appropriate structure for equivariant cohomology.
It is easy to write $S$ in a manifestly off-shell 
supersymmetric form by introducing a Lagrange multiplier, $\Lambda^{ij}.$

Notice that the terms in $S$ quartic in $X$ are not exactly in 
correspondence with a  dimensional reduction of a Yang-Mills action in 
$d+2$ dimensions.  One mismatch is the absence of the term  
$2\tr [X^{+},X^{-}]^{2},$ and the other mismatch is the coefficient of
the term $\tr [X^{+},X^{i}][X^{-},X_{i}].$  The first mismatch could 
be 
easily remedied by adding a term $2i\tr \ps -[X^{-},X^{+}]$ to $\Psi$ 
in eq.~\ref{gaugefermion}.  This term has the correct ghost number 
$-1$ to
be added to $\Psi,$  and leads merely to a small 
modification of the $[X_{i},\ps i]=0$ constraint. It also does not change the 
$b_{ij}$ equation of motion, and therefore is consistent with the
supersymmetry.  However, I do not add such a term because the 
action $S$ is easier to study as it stands, since $X^{\pm}$ appear 
linearly.  
The remaining mismatch is due to the fact that 
there are no fields $b^{-j}.$
The terms with anticommuting fields can be written 
as 
\begin{equation}
\tr\bigg( b_{[IJ}n_{K]}[X^{I},b^{JK}]\bigg),\qquad {\rm with}\  n_{J}\equiv 
\delta_{J+}, 
\label{anticomm}
\end{equation}  
 except for the coefficient of the term in $\Delta 
S.$ 

The condition $s^{2}=0$ is the analogue in the present case of the
BPS condition $\det\{Q_{\al},Q_{\be}\}=0.$  Thus, imposing the 
constraint  $[X^{+},X^{I}]=0,$ we find
\begin{equation}
[X_{j},[X^{j},X^{i}]] + {i\over 2}\{b^{+j},b^{i}{}_{j} - 
b^{+-}\delta^{i}{}_{j}\}=0,\qquad [X_{i},[X^{i},X^{-}]] - {i\over 4} 
\{b^{ij},b_{ij}\} =0.
\label{eom}
\end{equation}
Let us compare this set-up to the equations and constraints found by 
Nishino and Sezgin\cite{ns} in their construction of supersymmetric 
Yang-Mills equations in $10+2$ dimensions.  Consider a vector field 
$X^{I}$ and a 
positive chirality Majorana-Weyl fermion $\lambda$ 
in $10+2$ dimensions. Dimensionally reduced to $0+0$ 
dimensions\cite{me},  
their constraints\cite{ns} are  
\begin{equation}
[X^{+},X^{I}] = 0 = [X^{+},\lambda], \qquad \Gamma^{+}\lambda =0,
\label{constraints}
\end{equation}
and their bosonic equation  of motion is  
\begin{equation}
[X_{J},[X^{J},X_{[K}]]n_{L]} + {1\over 4} \{ 
\bar\lambda,\Gamma_{KL}\lambda\} = 0.
\label{nsequation}
\end{equation}
We see that supersymmetric solutions in the topological theory 
constructed above  
satisfy the same constraint of commuting with $X^{+},$ and that 
eq.~\ref{eom} has precisely the same bosonic terms as 
eq.~\ref{nsequation}.   
$\Gamma^{+}\lambda=0$ is the analogue of $b^{-j}=0,$ rather than 
$b^{+j}=0,$ since 
$\Gamma^{I}$ changes chirality.   I do not understand the 
exact mapping of anticommuting degrees of freedom, which  
depends on the signature of the metric.  However, 
Nishino\cite{nishino} has given an action underlying the Nishino-Sezgin
equations\cite{ns}, and this action has a fermion term $\tr\big( 
\bar\lambda\Gamma^{IJ}[X_{I},\lambda]n_{J}\big),$ which is 
similar to eq.~\ref{anticomm}.

This leads to the following conjecture: 
Precisely when $d=9+1,$ the topological model equations for 
supersymmetric solutions are a twisted form of the Nishino-Sezgin 
equations, with a twist related to SO(8) triality.  
If this conjecture holds, it may imply that an appropriate subset 
of topological observables in the `physical' theory ({\it i.e.} the theory with
anticommuting fields in spinor representations) can be computed 
reliably in perturbation theory.   
The topological theories in other 
dimensions probably  do not admit such twistings, and therefore are 
not of much physical interest.

Before going on to an analogous construction for a general algebra, I 
note that the weak coupling equations of motion for the topological model 
lead to 
formal solutions which take the form of singular differential 
forms\cite{note}
on the 
space of diagonal matrices that have poles when two (or 
more) of the $N$ points coincide in $\rd.$  These singular 
differential forms are  similar to those described in, for example,  
\cite{arnold}.    
 
\def\MM#1#2{S^{#1}{}_{{#2}}}
We see in the above topological construction for $\rd$ 
hints of the general structure of 
supersymmetric matrix models\cite{banks,wati,sen}.  
It is  natural to ask what the 
analogous construction would be for a general algebra.  
I discuss briefly the general structure of such spaces of algebra 
homomorphisms.  Suppose the 
structure constants of an algebra $A$ (resp. $B$) 
are $f_{\al}{}^{\be\ga}$ (resp. $g_{a}{}^{{bc}}$) with respect to a 
basis $\{s^{\al}\}$ (resp. $\{r^{a}\}$).  Then a linear map
$M:A\rightarrow B$ which takes $s^{\al}\mapsto \MM\al{a}r^{a}$ 
is an algebra homomorphism if 
\begin{equation}
\MM \al{a}\MM \be{b} g_{c}{}^{ab} = \MM \ga{c} f_{\ga}{}^{\al\be}.
\label{structure}
\end{equation}
I have 
replaced $x^{i}$ in the previous discussion with $s^{\al}$ in order 
to emphasize that this is a much more general setup, involving not just
some set of generators of the algebra, but {\it every} element of a basis 
for the vector space structure underlying the algebra.  
We ask what model results when we use eq.~\ref{structure} as 
the constraint equation, instead of $[X^{i},X^{j}]=0.$   
Of course, {\it if } $M$ is a homomorphism, it 
is determined completely by its action on generators of
$A,$ so one   could restrict the use of eq.~\ref{structure} only
to indices $\al,\be$ corresponding to such generators\cite{thesis}.  
Not
only is such a restricted subset of equations difficult to describe in 
an invariant way, but we have constructed above a measure 
on {\it arbitrary} linear maps, which is only peaked on the set of 
homomorphisms.  

When the algebra $B$ is $H_{N},$ eq.~\ref{structure} can  be written as 
$S^{\al}S^{\be}= 
S^{\ga}f_{\ga}{}^{\al\be},$
with matrix multiplication in $H_{N}$ automatically incorporating the
r\^ole of $g_{a}{}^{bc}$ in eq.~\ref{structure}.
We introduce a BRS operator $s,$ and 
ghost matrices with anticommuting entries $b^{+\al},$
such that 
\begin{equation}
sS^{\al}= b^{+\al},\qquad \ps{\al} = i[X^{+},S^{\al}],\qquad sX^{+}=0,
\label{sST}
\end{equation}
in analogy with eq.~\ref{sX}.  The number of 
matrices introduced here is infinite, even for $\rd.$
Thus, for $\rd,$ if we introduce a matrix $X^{1}$ for the 
co\"ordinate function $x^{1},$ we introduce an {\it independent} matrix 
for the function $(x^{1})^{2}.$ The principle is that we are 
integrating over {\it all} linear maps between the vector spaces 
underlying the algebras $A$ and $H_{N},$ with the action constraining 
the linear maps to be algebra homomorphisms.  Thus, it is logically 
necessary to map $(x^{1})^{2}$ to a matrix which is {\it a priori} 
independent of $(X^{1})^{2}.$     
These may be the analogues of massive modes in string theory.

The antighosts, $b^{\al\be},$ matrices with 
anticommuting entries, are paired with 
Lagrange multipliers, $\Lambda^{\al\be},$ which 
are matrices with commuting entries:
\begin{equation}
sb^{\al\be}= \Lambda^{\al\be},\qquad s\Lambda^{\al\be} = 
i[X^{+},b^{\al\be} ],\qquad sX^{+}=0.
\label{sbL}
\end{equation}
To proceed further, suppose that $A$ is equipped with an inner 
product---for example, if 
$A$ has a trace, we could use $\Tr_{A}(s^{\al}s^{\be})\equiv 
h^{\al\be}$ as the inverse of a metric.  The observables 
of the topological theory will be independent of this inner 
product, but possible twists to get physical fermionic fields in 
spinor representations will depend on the choice of inner product.  
We  use the inner product to raise and lower indices, and introduce
$X^{-}\equiv K,b^{+-}\equiv L,$ as in eq.~\ref{sb}.

The  gauge fermion 
\begin{equation}
\Psi_{A} \equiv 2\tr\left[b_{\al\be}\left(S^{\al}S^{\be}-
S^{\ga}f_{\ga}{}^{\al\be} - {1\over 2}\Lambda^{\al\be}\right) +
i\ps \al[X^{-},S_{\al}]\right]
\label{fermiontwo}
\end{equation}
leads to an action, $S_{A}\equiv s\Psi_{A},$ 
that after integrating over $\Lambda,$ has terms 
quadratic, cubic and quartic in $S^{\al},$  somewhat reminiscent of 
string field theory.  
It is instructive  to see how the $[X^{i},X^{j}]^{2}$
term arises for $\rd$ by integrating out the mode corresponding to
$x^{i}x^{j}\equiv x^{j}x^{i}.$  Thus,   the Yang-Mills form
of matrix model actions for $\rd$ is  an artifact of the fact that we
are considering an algebra $A\equiv \cd$ that happens to be commutative.
The action  $S_{A}$ is the appropriate general form---it is 
important to note that minimal assumptions were made about the 
properties of
$A.$  (In particular, it may be of interest to take $A$ to be a 
Poisson algebra, possibly associated with a symplectic structure on a 
manifold, in which case $S^{\al}S^{\be}$ is replaced by 
$[S^{\al},S^{\be}]$ in eq.~\ref{fermiontwo}, with all else unchanged.) 
Integrating 
out appropriate $S^{\al}$ will reduce the functional integral to a
more nonlinear form, expressed in terms of generators and relations 
for the algebra, $A.$
Note especially 
that there seems to be a clear distinction between $X^{\pm}$ and 
$S^{\al}$ in $S_{A}.$

Even if the conjecture stated earlier  can be proved, what does all this 
have to do with physics?  First, we need to understand 
what restrictions the existence of a twisted spinorial interpretation for 
anticommuting fields places on the pairs $A,h_{\al\be}.$ A possible 
line of attack is to take the bosonic terms of $S_{A}$ and work out 
conditions for the existence of a
supersymmetric extension.  It may be of interest to consider the work of 
Baulieu, Green and Rabinovici\cite{green} in this context.   
{\it If} one could obtain such an understanding, I  suggest the following 
interpretation:  Physical observations involve a very large number of
`events'.  When we consider   amplitudes for  large numbers of 
events, we find that there are saddlepoints in the matrix integral 
that can be interpreted as
the propagation of particles or  extended objects from a spacetime perspective.    
Nevertheless, the underlying theory may just be a `theory of events', 
with  conditional  amplitudes calculated from matrix integrals.  What 
is interesting is that the matrix integral automatically includes 
matrices which na\"\i vely  correspond to two additional 
dimensions, which may perhaps be related to  F theory 
compactifications\cite{vafa}.  It may be interesting to consider a 
topological matrix model for configurations of $N$ lines (or other 
extended objects), as well.
 

I am  grateful to R. von Unge for valuable comments.
This work was supported in part by NSF grant PHY96-00258.


\begin{references}
\bibitem{note} The anticommuting fields do  {\it not} vanish for 
such solutions, so these solutions should not be confused with the
supersymmetric configurations discussed above.
\end{references}
\end{document}